\documentclass[superscriptaddress,aps,preprintnumbers,amsmath,showpacs,amssymb,nofootinbib,reprint]{revtex4-1}
\usepackage{bm, color} 
\usepackage{amssymb,amsfonts,slashed,amsthm,amsmath,graphicx, soul}
\usepackage{subfigure}
\usepackage{hyperref}
\usepackage{tikz,xcolor}

\begin{document}

%%%%%%%%%%%%%%%%%%%%%%%%%%%%%%%%%%%%%%%%%%%

\newcommand{\vev}[1]{ \left\langle {#1} \right\rangle }
\newcommand{\bra}[1]{ \langle {#1} | }
\newcommand{\ket}[1]{ | {#1} \rangle }
\newcommand{\eV}{ \ {\rm eV} }
\newcommand{\KeV}{ \ {\rm keV} }
\newcommand{\MeV}{\  {\rm MeV} }
\newcommand{\GeV}{\  {\rm GeV} }
\newcommand{\TeV}{\  {\rm TeV} }
\newcommand{\1}{\mbox{1}\hspace{-0.25em}\mbox{l}}
\newcommand{\Red}[1]{{\color{red} {#1}}}

\newcommand{\lmk}{\left(}  
\newcommand{\rmk}{\right)}
\newcommand{\lkk}{\left[}  
\newcommand{\rkk}{\right]}
\newcommand{\lhk}{\left \{ }  
\newcommand{\rhk}{\right \} }
\newcommand{\del}{\partial}  
\newcommand{\la}{\left\langle} 
\newcommand{\ra}{\right\rangle}
\newcommand{\half}{\frac{1}{2}}

\newcommand{\bea}{\begin{array}}
\newcommand{\eea}{\end{array}}
\newcommand{\beq}{\begin{eqnarray}}
\newcommand{\eeq}{\end{eqnarray}}
\newcommand{\eq}[1]{Eq.~(\ref{#1})}

\newcommand{\dd}{\mathrm{d}}
\newcommand{\Mpl}{M_{\rm Pl}}
\newcommand{\mg}{m_{3/2}}
\newcommand{\abs}[1]{\left\vert {#1} \right\vert}
\newcommand{\mphi}{m_{\phi}}
\newcommand{\Hz}{\ {\rm Hz}}
\newcommand{\for}{\quad \text{for }}
\newcommand{\Min}{\text{Min}}
\newcommand{\Max}{\text{Max}}
\newcommand{\Kahler}{K\"{a}hler }
\newcommand{\cphi}{\varphi}
\newcommand{\Tr}{\text{Tr}}
\newcommand{\diag}{{\rm diag}}

\newcommand{\SUf}{SU(3)_{\rm f}}
\newcommand{\Upq}{U(1)_{\rm PQ}}
\newcommand{\Zpq}{Z^{\rm PQ}_3}
\newcommand{\Cpq}{C_{\rm PQ}}
\newcommand{\ubar}{u^c}
\newcommand{\dbar}{d^c}
\newcommand{\ebar}{e^c}
\newcommand{\nubar}{\nu^c}
\newcommand{\Ndw}{N_{\rm DW}}
\newcommand{\Fpq}{F_{\rm PQ}}
\newcommand{\fpq}{v_{\rm PQ}}
\newcommand{\Br}{{\rm Br}}
\newcommand{\Lag}{\mathcal{L}}
\newcommand{\Lqcd}{\Lambda_{\rm QCD}}

\newcommand{\ji}{j_{\rm inf}} 
\newcommand{\jb}{j_{B-L}} 
\newcommand{\M}{M} 
\newcommand{\im}{{\rm Im} }
\newcommand{\re}{{\rm Re} }

\def\lrf#1#2{ \left(\frac{#1}{#2}\right)}
\def\lrfp#1#2#3{ \left(\frac{#1}{#2} \right)^{#3}}
\def\lrp#1#2{\left( #1 \right)^{#2}}
\def\REF#1{Ref.~\cite{#1}}
\def\SEC#1{Sec.~\ref{#1}}
\def\FIG#1{Fig.~\ref{#1}}
\def\EQ#1{Eq.~(\ref{#1})}
\def\EQS#1{Eqs.~(\ref{#1})}
\def\TEV#1{10^{#1}{\rm\,TeV}}
\def\GEV#1{10^{#1}{\rm\,GeV}}
\def\MEV#1{10^{#1}{\rm\,MeV}}
\def\KEV#1{10^{#1}{\rm\,keV}}
\def\blue#1{\textcolor{blue}{#1}}
\def\red#1{\textcolor{red}{#1}}

\newcommand{\eff}{\Delta N_{\rm eff}}
\newcommand{\neff}{\Delta N_{\rm eff}}
\newcommand{\cc}{\Lambda}
\newcommand{\Mpc}{\ {\rm Mpc}}
\newcommand{\Msolar}{M_\cdot}

\def\sn#1{\textcolor{blue}{#1}}
\def\SN#1{\textcolor{blue}{[{\bf SN:} #1]}}

% %######################
%\preprint{TU-xxxx}
% %######################

\title{
Phase transition dynamics of Abelian-Higgs vortices in the oscillating axion background
}

\author{Naoya Kitajima}
\affiliation{Frontier Research Institute for Interdisciplinary Sciences, Tohoku University, \\
6-3 Azaaoba, Aramaki, Aoba-ku, Sendai 980-8578, Japan}
\affiliation{Department of Physics, Tohoku University, \\
6-3 Azaaoba, Aramaki, Aoba-ku, Sendai 980-8578, Japan}

\author{Shota Nakagawa}
\affiliation{Tsung-Dao Lee Institute, Shanghai Jiao Tong University, \\
No.~1 Lisuo Road, Pudong New Area, Shanghai 201210, China}
\affiliation{School of Physics and Astronomy, Shanghai Jiao Tong University, \\
800 Dongchuan Road, Shanghai 200240, China}

\begin{abstract}
We investigate the effect of an oscillating axion background on thermally induced vortex-antivortex production in the Abelian-Higgs model.
Through its coupling to the gauge field, the axion sources an electromagnetic response in the presence of vortex magnetic flux, modifying the dynamics of the vortices. 
We study this system using Langevin lattice simulations with dissipation and stochastic noise.
It is shown that the phase transition dynamics of Abelian-Higgs vortices, namely the pair creation of vortex and antivortex under thermal fluctuations, can be significantly modified in the oscillating axion background.
We find that the axion background shifts the rapid increase in the vortex density to lower temperature. 
In particular, this transition temperature becomes smaller as the oscillation frequency of the axion increases. 
The temperature dependence of the correlation length is also modified.
These features provide a new experimental approach for the axion dark matter using superconducting devices.
\end{abstract}

\maketitle
\flushbottom

\section{Introduction}

Axion is a well-motivated candidate for dark matter in our Universe.
Due to the shift symmetry, light axions are theoretically predicted as the QCD axion \cite{Peccei:1977hh,Peccei:1977ur,Weinberg:1977ma,Wilczek:1977pj} and also as axion-like particles e.g. in string theory \cite{Svrcek:2006yi,Arvanitaki:2009fg,Cicoli:2012sz}.
When the axion mass becomes comparable to the Hubble scale of the Universe, the axion begins to oscillate around the potential minimum and behaves as a coherent oscillation \cite{Preskill:1982cy,Abbott:1982af,Dine:1982ah}.
In addition, the axion can be created from the decay of topological defects, such as cosmic strings and domain walls.
These contributions can explain the observed dark matter abundance in broad parameter spaces.
See \cite{Sikivie:2006ni,Kawasaki:2013ae,Marsh:2015xka,DiLuzio:2020wdo,Choi:2020rgn,OHare:2024nmr} for reviews.
If the axion has the Chern-Simons coupling to the electromagnetic field, 
the electromagnetic response provides a means of probing the axion background (see e.g. \cite{Irastorza:2018dyq} for review of experimental axion probes). 
In particular, in an external magnetic field, the time dependence of the axion field acts as a source for an electric field, which can leave characteristic signatures in experimental devices.

Superconducting materials are a very rich stage with electromagnetic reaction. 
The phase dynamics of superconductors is effectively described by the Abelian-Higgs model, or Ginzburg-Landau theory. 
One of the interesting features is a static solution known as a vortex or cosmic string that consists of quantized magnetic flux with an integer winding number \cite{Abrikosov:1957wnz,Nielsen:1973cs}.
In our previous work \cite{Kitajima:2025nml}, we studied the Abelian-Higgs model with the axion interacting with the gauge field.
We have shown that the vortex configuration and its dynamics can be significantly modified in the presence of the oscillating axion background.
The oscillating axion induces electric and magnetic fields by coupling with the vortex magnetic flux. 
The induced fields modify inter-vortex interactions and can produce either attraction or repulsion even at the critical coupling, where the static force between vortices vanishes in the absence of the axion.
In our previous work, we only presented what happens numerically but did not propose a more reliable detection method of the axion dark matter.
Our goal of the present paper is to investigate this system furthermore and identify observables that can be detected in future experiments.

To probe the axion through feeble coupling, 
we need to predict observable signals in experiments.
In this work, we focus on the phase transition dynamics associated with the collective behavior of the Abelian-Higgs vortices.
Thermally induced vortices and antivortices play a central role in the phase transition dynamics of two-dimensional systems.
In a system with global U(1) symmetry, the Berezinskii-Kosterlitz-Thouless (BKT) transition is driven by the unbinding of vortex-antivortex pairs~\cite{Berezinskii:1972fet,Kosterlitz:1973xp,Kosterlitz:1974nba}.
At low temperatures, vortices appear predominantly as bound pairs, 
whereas unbound vortices proliferate at higher temperatures.
As a result, the correlation length suddenly changes at some transition temperature \cite{Nelson:1977zz}.
Although the gauge field can significantly modify this transition dynamics of global vortices,
superconducting systems also have similar transition phenomena \cite{PhysRevLett.42.1165,halperin1979resistive,PhysRevB.28.5075,PhysRevLett.131.186002}.

In this paper, we investigate thermally induced vortex-antivortex production in the Abelian-Higgs model in the presence of an oscillating axion background. 
We describe thermal contact with an external thermal bath using Langevin field equations with the fluctuation-dissipation theorem and perform numerical lattice simulations.
This framework allows us to explore how the axion-induced electromagnetic response affects the phase transition dynamics.

We find that the oscillating axion background shifts the transition temperature at which the vortex density rises rapidly.
This temperature decreases as the axion frequency increases over the range explored in our simulations.
Moreover, we examine the correlation functions to observe a collective phase dynamics of vortices that can deviate from a conventional BKT-like transition.
In our simulations, we find that the correlation length drops at the transition temperature.
These results motivate further investigation of superconducting systems as possible probes of the axion dark matter.

The remainder of this paper is organized as follows.
In \SEC{sec:AH_model}, we introduce the Abelian-Higgs model coupled to the axion background.
To realize the system with thermal contact,
we take the Langevin dynamics approach in \SEC{sec:Langevin}.
In \SEC{sec:results}, we present results of numerical lattice Langevin simulations, and \SEC{sec:discussion} is devoted to discussion and conclusion.
Appendix \ref{app:interaction} shows the interaction between vortices and antivortices in the oscillating axion background without thermal noise.
In Appendix \ref{app:Gauss}, we show how the Gauss constraint is maintained in the presence of an external thermal contact.

\section{Abelian-Higgs model with axion
\label{sec:AH_model}}

Let us start with an Abelian-Higgs model or a Ginzburg-Landau (GL) model, where a complex scalar (Higgs) field or GL order parameter, $\Phi$, describes phase dynamics in superconducting materials.
Moreover, we introduce an axion field, $a$, interacting with gauge fields, $A_\mu$, based on the same setup as in the previous paper \cite{Kitajima:2025nml} (see also \cite{Kitajima:2021bjq,Nakagawa:2022knn,Kitajima:2025jct} for the Abelian-Higgs model in a dark sector in the cosmological context). 
The Lagrangian is given by
\begin{align}
\mathcal{L} &= (D_\mu \Phi)^\dagger D^\mu \Phi -V_\Phi(\Phi) - \frac{1}{4} F_{\mu\nu}F^{\mu\nu}\nonumber\\
&+ \frac{1}{2}\del_\mu a\del^\mu a -V_a(a) -\frac{g_{a\gamma}}{4} a F_{\rm \mu\nu}\tilde{F}^{\mu\nu}.
\label{Lag}
\end{align}
Here $D_\mu = \partial_\mu - ieA_\mu$ is the gauge covariant derivative, $F_{\mu\nu} = \partial_\mu A_\nu - \partial_\nu A_\mu$ is the field strength tensor, $\tilde{F}^{\mu\nu} = \epsilon^{\mu\nu\rho\sigma} F_{\rho\sigma}/2$ is its dual, 
and the axion has the Chern-Simons coupling with a dimensionful coupling constant $g_{a\gamma}=c_\gamma\alpha/(\pi f_a)$, where $f_a$ denotes the axion decay constant, $c_\gamma$ the anomaly coefficient for the U$(1)$ gauge symmetry, and $\alpha\equiv e^2/(4\pi)$.

We consider a Mexican-hat potential for the Higgs field,
\begin{align} \label{eq:potential}
    V_\Phi(\Phi) = \frac{\lambda}{4} (|\Phi|^2 - v^2)^2,
\end{align}
with $\lambda$ being the self-coupling constant and $v$ the vacuum expectation value.
The axion potential $V_a$ drives the axion fields to oscillate at the minimum, leading to the misalignment production \cite{Preskill:1982cy,Abbott:1982af,Dine:1982ah} and the creation of string-wall structure which decays to the axions \cite{Davis:1986xc,Lyth:1991bb} (see also, e.g., \cite{Hiramatsu:2010yn,Hiramatsu:2010yu,Hiramatsu:2012gg,Hiramatsu:2012sc,Kawasaki:2014sqa}).
As long as it is light enough, the axion behaves as a coherent oscillation in a realistic setup of table-top experiment.
Throughout the paper, we focus on the case in which the axion is a light dark matter component in the Universe. 
Specifically, we assume that the axion is a spatially homogeneous coherent oscillation in the following form,
\begin{align} \label{eq:axion_osc}
a(t) = \mathcal{A} \cos(\omega_a t + \delta),
\end{align}
where $\mathcal{A}$ and $\omega_a( \approx m_a)$ are the amplitude and frequency of the axion oscillation respectively, and $\delta$ is the initial phase. 
This simplification is justified when we consider the physical system of superconducting devices whose typical size is well within the de Broglie wavelength of the axion.
In what follows, we set $\delta=0$ without loss of generality, and $g_{a\gamma}$ is absorbed by the definition of $\mathcal{A}$.

It is well-known that the potential (\ref{eq:potential}) admits a static solution, the so-called Abrikosov-Nielsen-Olesen (ANO) vortex \cite{Abrikosov:1957wnz,Nielsen:1973cs}.
When the U$(1)$ gauge symmetry is spontaneously broken, vortices (or cosmic strings) are copiously produced.
The mass of the Higgs and gauge fields are given by, respectively, $m_\Phi = \sqrt{\lambda}v$ and $m_A = \sqrt{2}ev$ in the symmetry-broken phase, while the symmetry is restored inside the vortex and the gauge fields are massless there, allowing the existence of stable magnetic flux.
The interplay between the Higgs and gauge field configurations changes the free energy which induces the interaction among (anti) vortices \cite{Jacobs:1978ch,Taubes:1979tm,Ruback:1988ba,Moriarty:1988fx,Myers:1991yh,Samols:1991ne,Rebbi:1991wk,Speight:1996px,Eto:2022hyt,Fujikura:2023lil,Hirose:2024wuk}.
In the absence of the axion background, the physical property is determined by the mass ratio
\begin{align}
\beta\equiv \frac{m_\Phi^2}{m_A^2} = \frac{\lambda}{2e^2}.
\end{align}
In other words, the interaction is relevant to the ratio between the correlation length (scalar core size) and the magnetic penetration length (vector core size).
Namely, $\beta < 1$ ($\beta > 1$), corresponding to the type-I (II) superconductor, results in the attractive (repulsive) interaction between two vortices.
In the special case with $\beta = 1$, known as the critical coupling or Bogomol’nyi-Prasad-Sommerfield (BPS) limit \cite{Bogomolny:1975de,Prasad:1975kr}, no force is exerted between two vortices.

In the previous paper \cite{Kitajima:2025nml}, we showed that the magnetic field resonantly enhances electromagnetic fields inside the vortex core like a cylindrical cavity and alters the interaction among the vortices due to the axion-photon coupling.
Specifically, the interaction turns either attractive or repulsive even in the case with the critical coupling $(\beta=1)$.
In addition, the vortex-antivortex interaction can also be significantly modified in the presence of the axion background, as shown in Appendix \ref{app:interaction}.
Thus, it is natural to expect that the corrective dynamics of vortices and antivortices is altered.
In the next sections, we explore the dynamics of vortex-antivortex pair creations in the oscillating axion background.

\section{Langevin field equations
\label{sec:Langevin}}

A realistic superconducting device is a finite-temperature system with thermal fluctuations originating from contact with an external thermal bath.
Here we consider the Abelian-Higgs model with thermal fluctuations to explore the possibility that the oscillating axion background leaves its footprint on the vortex dynamics in superconductors.

The vortex configuration is an essential ingredient for the phase dynamics in condensed-matter systems.
It is well-known that there exists a quasi-ordered phase at a low temperature in two-dimensional space, despite the Mermin-Wagner theorem \cite{Mermin:1966fe}. 
The quasi-ordered phase transitions to a disordered phase as the temperature increases, which is called the Berezinskii-Kosterlitz-Thouless (BKT) transition \cite{Berezinskii:1972fet,Kosterlitz:1973xp,Kosterlitz:1974nba} (see \cite{Minnhagen:1987zz,Tong:SFT,Kosterlitz_2016,Drouin-Touchette:2022ydh} for review papers).
At a low temperature, vortices and antivortices are bound into pairs, which mimics the ordered phase, while these pairs come to dissociate at a high temperature, leading to disorder.
The transition temperature is well estimated under a global U(1) symmetry by comparing the energy to create a vortex and the entropy, corresponding to the XY model realized by superfluid systems \cite{Nelson:1977zz}.
In superconducting systems, where the order parameter is coupled to the electromagnetic gauge field, analogous vortex-antivortex
unbinding behavior is also expected and has been studied theoretically
\cite{PhysRevLett.42.1165,halperin1979resistive}
and experimentally
\cite{PhysRevB.28.5075,PhysRevLett.131.186002}.
In this paper, we explore the pair creation of vortices and antivortices and their dissociation in the axion background.

To describe the dynamics of vortices with thermal fluctuations,
we adopt the Langevin dynamics approach in classical field theory.
The thermal interaction is approximately described by stochastic processes in which the microscopic physics is coarse-grained, so that a background thermal bath with temperature $T$ exchanges energy with our system. 
Adding the thermal fluctuation to the Euler-Lagrange equations derived from the Lagrangian (\ref{Lag}), 
we have the system of Langevin equations for the canonical momenta, $\Pi = \dot{\Phi}$ and $E_i = \dot{A}_i$, together with the Gauss constraint (see e.g. \cite{Yates:1998kx,Bettencourt:2001ru}),
\begin{align}
&\dot{\Pi} =  D_i D_i \Phi -\frac{\del V_\Phi}{\del\Phi^*} -\eta\Pi +\xi_\Pi,
\label{Higgs_Lange}\\
&\dot{E}_i = -\nabla^2A_i +\del_i\del_jA_j - 2e{\rm Im}[\Phi^* D_i\Phi] \nonumber \\ & \qquad -\eta E_i +\xi_{E_i} +g_{a\gamma}\dot{a} B_i,
\label{Ampere_Lange} \\
&\del_i E_i = -2e{\rm{Im}}[\Phi^*\Pi],
\label{Gauss_Lange}
\end{align}
where we adopt the temporal gauge, $A_0=0$, $B_i \equiv \epsilon_{ijk} \partial_j A_k$, and the overdot represents the time derivative.
Note that we have dropped spatial derivatives of the axion field.
As $\Phi = (\phi_1+i\phi_2)/\sqrt{2}$, we parameterize the complex-valued $\Pi$ and $\xi_\Pi$ as $\Pi = (\pi_1+i\pi_2)/\sqrt{2}$ and $\xi_\Pi = (\xi_{\pi_1}+i\xi_{\pi_2})/\sqrt{2}$.
$\eta$ represents the viscosity coefficient, and $\xi_\alpha$ with $\alpha = \pi_i,E_i$ is the Gaussian white noise, 
satisfying the fluctuation-dissipation relation,
\begin{align}
\langle \xi_\alpha(t,\vec x) \xi_\beta(t',\vec x') \rangle = 2\eta T \delta_{\alpha\beta}\delta(t-t')\delta^2(\vec x-\vec x') \ .
\label{FDT}
\end{align}
Here $\vec x =(x,y)$ is the transverse vector, and we assume that the shift symmetry in the $z$-direction, corresponding to the direction of the quantized magnetic flux, in other words, it is essentially the $(2+1)$-dimensional spacetime. It is valid when one considers thin superconducting films.

In general, the interaction with the external thermal bath would easily spoil the Gauss constraint (\ref{Gauss_Lange}).
This can be avoided by projecting the random noise onto the constraint-preserving surface.
The technical details are given in Appendix \ref{app:Gauss}.

\section{Numerical results
\label{sec:results}}

\begin{figure*}[tp]
\centering
\subfigure{\includegraphics[width = 8.5cm, clip]{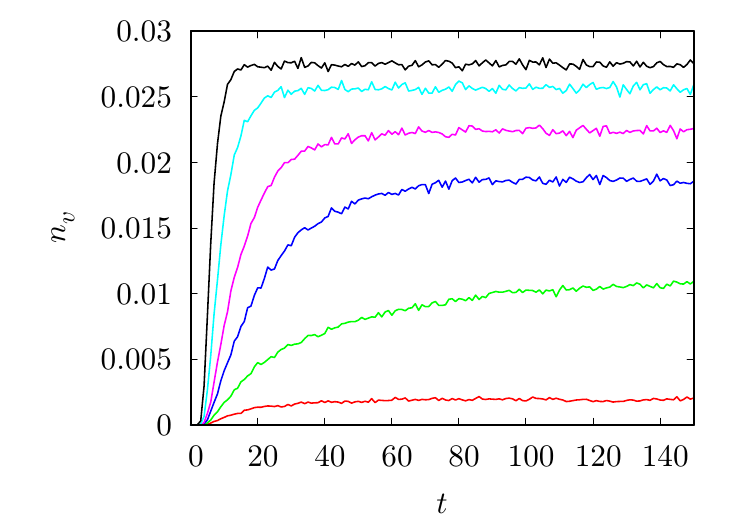}}
\label{subfig:timeseries_axion_1}
\subfigure{\includegraphics[width = 8.5cm, clip]{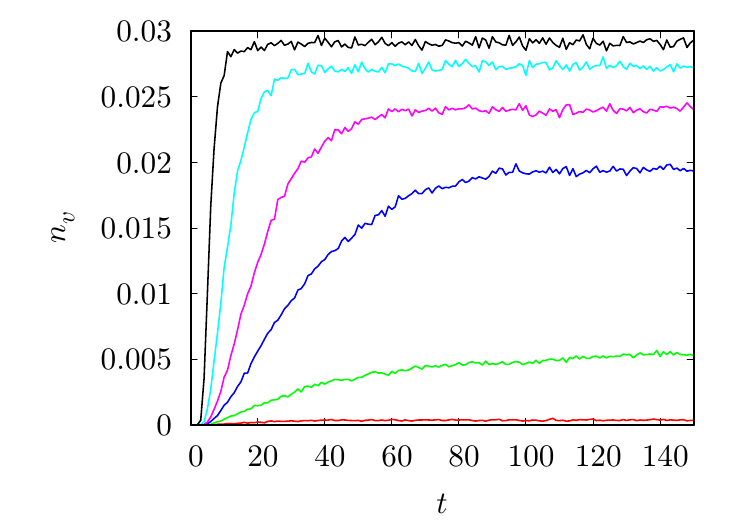}}
\label{subfig:timeseries_axion_3}
\caption{
Time evolution of the vortex number density. We set $\omega_a = 1$ ($2$) and $T = 0.12,\,0.13,\,0.14,\,0.15,\,0.17,\,0.2$ ($0.1,\,0.11,\,0.13,\,0.14,\,0.15,\,0.2$) from bottom to top in the left (right) panel.
}
\label{fig:timeseries}
\end{figure*}

We have performed numerical simulations to solve the Langevin equations, \EQS{Higgs_Lange},(\ref{Ampere_Lange}), with the oscillating axion background \EQ{eq:axion_osc} based on the lattice gauge formulation \cite{Moriarty:1988fx}.
In our simulations, we set the grid number $N^2 = 512^2$, the lattice spacing $\delta x = 0.25/v$ (the boxsize $L = 128/v$) and the stepsize for the time evolution $0.04 \delta x$. As a consistency check, we also performed simulations with different grid points and boxsizes, but we obtained the same results.
We start the simulation from the completely ordered phase by setting $\Phi = v$ (i.e. ${\rm Arg}(\Phi) = 0$), $\dot{\Phi} = 0$, $A_i = 0$ and $E_i = 0$.
The derivative is discretized by the second order finite difference and we update fields using the second order leap-frog method (for the deterministic part).
We fix the dissipation coefficient $\eta = 1$ and vary the temperature.
We performed 30 simulations for each temperature with different sets of random noise.

The number of vortices and antivortices is counted using the winding number based on the gauge-invariant phase differences around each lattice plaquette \cite{Kajantie:1998bg}. The size of the vortex core is given by $\delta_v = (\sqrt{\lambda/2}v)^{-1}$ and a single vortex contains $(\delta_v/\delta x)^2$ lattice points.  
Thus, when we count the number of isolated (anti)vortices, we drop the vortex-antivortex pair with the separation smaller than this core size.

\begin{figure*}[tp]
\centering
\subfigure{\includegraphics[width = 8.5cm, clip]{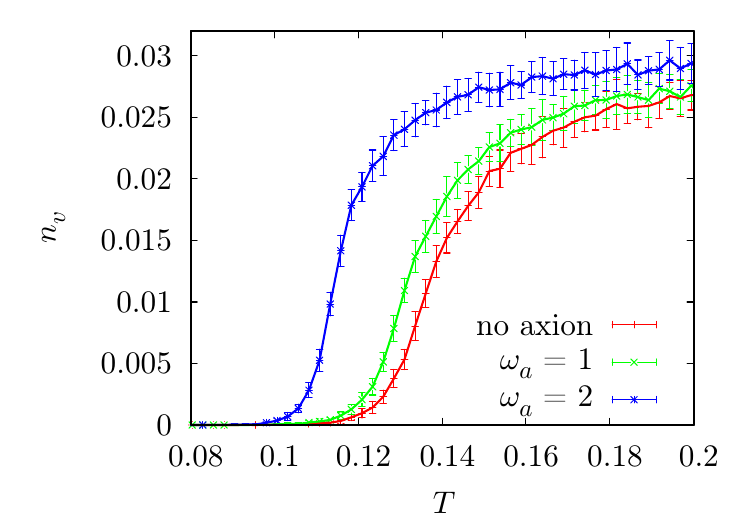}}
\label{subfig:vortex_T}
\subfigure{\includegraphics[width = 8.5cm, clip]{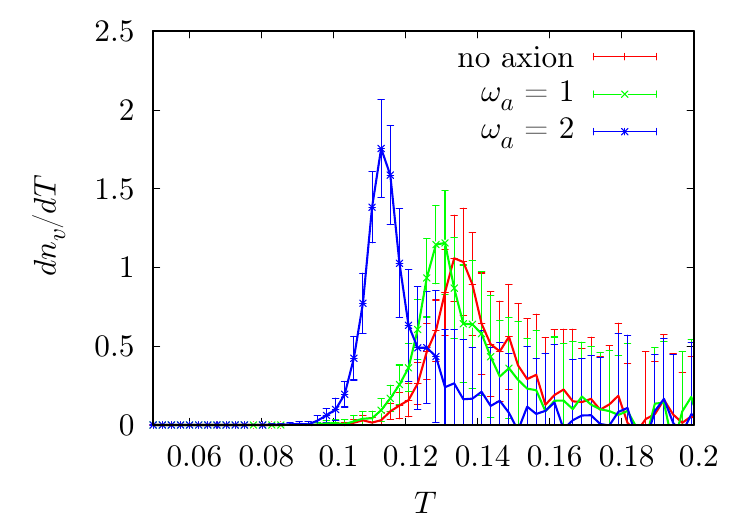}}
\label{subfig:vortex_T_log}
\caption{
Temperature dependence of the vortex number density at the final time (left) and its temperature derivative (right). The error bar corresponds to the 1-$\sigma$ error. The red, green and blue lines correspond respectively to the case without axion, $\omega_a = 1$ and $\omega_a = 2$.
}
\label{fig:vortex_T}
\end{figure*}

\begin{figure*}[tp]
\centering
\includegraphics[width = 8.5cm, clip]{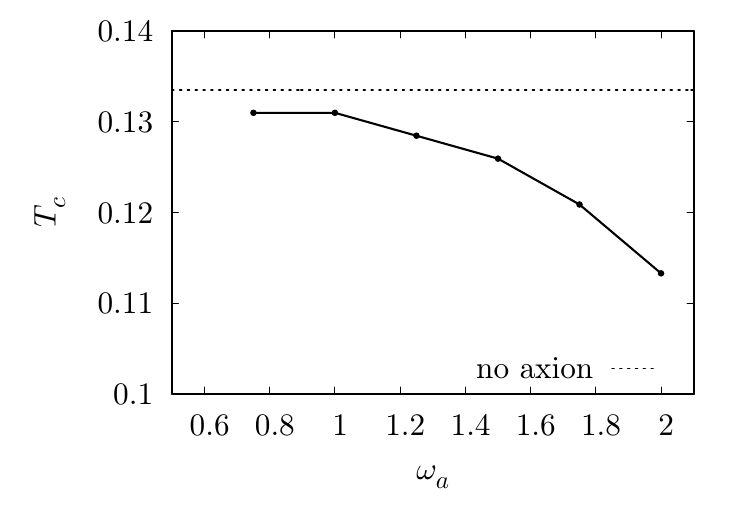}
\caption{
Transition temperature as a function of the axion frequency.
The horizontal dotted line corresponds to the case without axion.
}
\label{fig:Tc}
\end{figure*}

In what follows, the physical quantities are normalized by the Higgs expectation value, $v$, and we fix $\mathcal{A} = 1$ and $\beta=1$ (BPS limit).
Fig.~\ref{fig:timeseries} shows the time evolution of the vortex (and antivortex) number density, $n_v$, averaged over 30 simulations. It shows that the vortex density reaches saturation within the simulation time, which confirms that the configuration becomes thermal equilibrium. 
In Fig.~\ref{fig:vortex_T}, the left panel shows the temperature-dependence of the vortex number density at the final time of the simulation ($t = 150$) and the right panel shows its temperature derivative. The error bar shows the 1-$\sigma$ error.
Let us define the transition temperature by the one that maximizes $dn_v/dT$.
This figure clearly shows that the transition temperature is shifted in the presence of the background axion oscillation.
Note also that the value of $dn_v/dT$ increases as the axion frequency increases.
Fig.~\ref{fig:Tc} shows the transition temperature in terms of the axion frequency.\footnote{We note that the transition temperature should be lower than the critical temperature of superconductors corresponding to $v$ in our setup.
Otherwise, the proliferation of (anti)vortices would not occur while keeping superconductivity.}
This figure indicates that the transition temperature monotonically decreases as the axion frequency increases.

\begin{figure*}[tp]
\centering
\subfigure{\includegraphics[width = 8.5cm, clip]{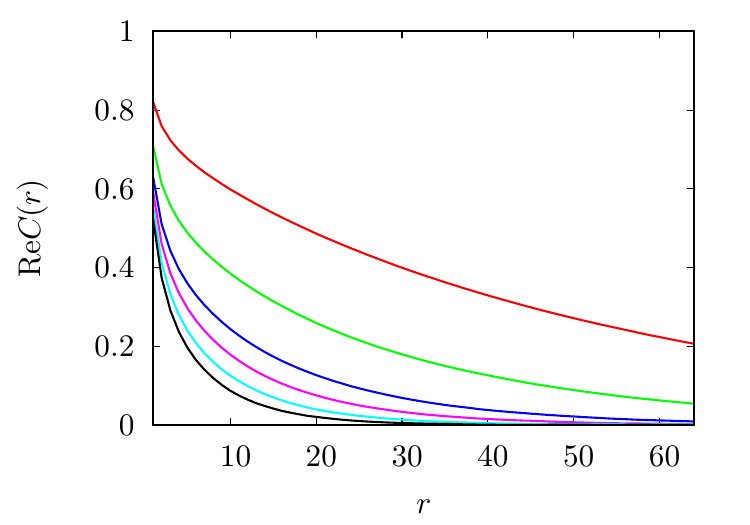}}
\label{subfig:phase_correlation_omega_1}
\subfigure{\includegraphics[width = 8.5cm, clip]{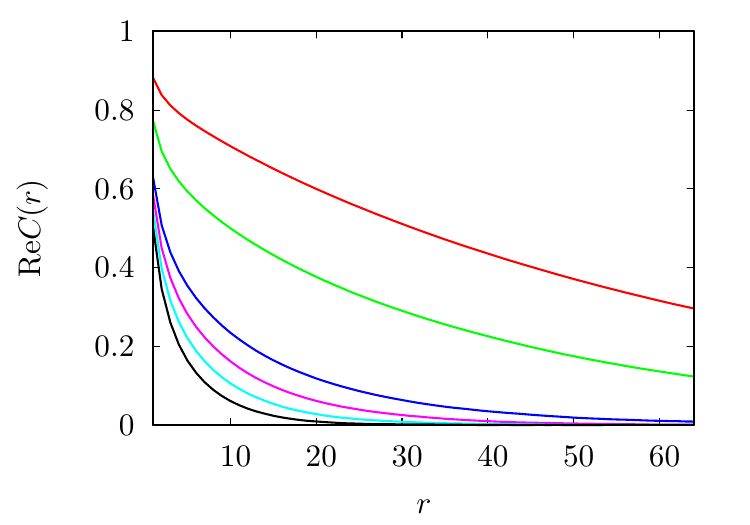}}
\label{subfig:phase_correlation_omega_2}
\caption{
Real part of the phase correlation function.
We set $\omega_a = 1$ ($2$) and $T = 0.12,\,0.13,\,0.14,\,0.15,\,0.17,\,0.2$ ($0.1,\,0.11,\,0.13,\,0.14,\,0.15,\,0.2$) from top to bottom in the left (right) panel.
}
\label{fig:phase_correlation}
\end{figure*}

\begin{figure}[tp]
\centering
\includegraphics[width = 8.5cm, clip]{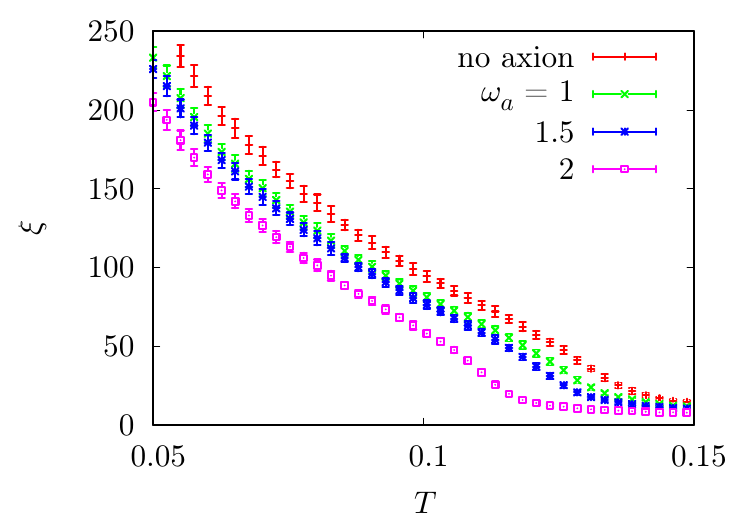}
\caption{
The correlation length as a function of the temperature.
}
\label{fig:correlation_length}
\end{figure}

We also calculated the phase correlation function and the correlation length.
The phase correlation function is defined in a gauge-invariant way as follows \cite{Gasenzer:2013era},
\begin{align}
C(r) = \langle \varphi^*(t,\vec{x}) U(t,\vec{x},\vec{y}) \varphi(t,\vec{y}) \rangle,
\end{align}
where the angular bracket denotes the ensemble average and
\begin{align}
\varphi(t,\vec{x}) &= \frac{\Phi(t,\vec{x})}{|\Phi(t,\vec{x})|},\\
U(t,\vec{x},\vec{y}) &= \exp \left[ -ie \int^y_x \vec{A}(t,\vec{x}') \cdot d\vec{x}' \right].
\end{align}
Practically, we compute the correlation function by sampling the above product along the $x$ and $y$-axes in the lattice space.
We can then obtain the correlation length by fitting the correlation function with an exponentially decaying function \cite{jelic2011quench},
\begin{align}
C(r) \propto e^{-r/\xi}.
\end{align}
Fig.~\ref{fig:phase_correlation} shows the real part of the phase correlation function at the final time of the simulation. Different curves correspond to different temperatures.
We actually found that the correlation function can be well-fitted by the exponentially decaying function, which enables us to compute the correlation length. Fig.~\ref{fig:correlation_length} shows the correlation length as a function of the temperature.
This figure illustrates that the correlation length becomes shorter in the presence of the background axion oscillation. 
In particular, the temperature dependence changes significantly in the case with $\omega_a = 2$.
It can give us an independent measurement to probe the axion background.

\section{Discussion
\label{sec:discussion}}

In this paper, we have shown that the oscillating axion background significantly affects the dynamics of the vortex-antivortex pair creation under thermal fluctuations.
Namely, the transition temperature can be significantly shifted.
Our results show that the transition temperature decreases with the increase of the axion frequency. This is because the electric field is induced with the source term proportional to $\dot{a}$. Thus the modification may depend on the combination $\omega_a\mathcal{A}$. It should be confirmed with more detailed simulations. It is left for the future work.

In our analysis, we assume that the vortex core is in the vacuum state, whereas it should be a normal conducting state.
Since the induced electromagnetic fields are damped by the finite conductivity, the evolution of the vortex configuration may be affected inside the core.
In particular, the resonant enhancement studied in the previous paper \cite{Kitajima:2025nml} could be affected significantly.
However, the vortex interaction or the collective behavior of vortices is mainly determined by the configuration in regions far from the core.\footnote{
In \FIG{fig:Tc}, no peculiar change appears at the resonant frequency, $\omega_a\simeq 1.25$, estimated in the previous paper \cite{Kitajima:2025nml}.
This may be because the oscillation period of the axion is longer than the timescale of the vortex-antivortex pair creation/annihilation,
$\omega_a \lesssim 2\pi(\eta T_c)^{1/2}$.
In realistic cases for a light axion, e.g. $v\sim {\rm meV}$ and $\omega_a\sim1\mu {\rm eV}$, this bound can be typically satisfied.}
Therefore, we expect that the results presented in this paper would not be significantly modified.
The conclusion will be clarified in a separate paper.

\vspace{2mm}
{\bf Acknowledgments.--}
We thank Satoshi Kashiwaya and Yasunori Mawatari for fruitful discussions.
This work was supported by JSPS KAKENHI (Grant Number 26K00695 [NK]).
This work used computational resources of Fugaku supercomputer, provided by RIKEN Center for Computational Sciences, through the HPCI System Research Project (Project ID: hp260117).

\appendix

\section{Vortex-antivortex interaction
\label{app:interaction}}

\begin{figure}[tp]
\centering
\includegraphics[width = 8.5cm, clip]{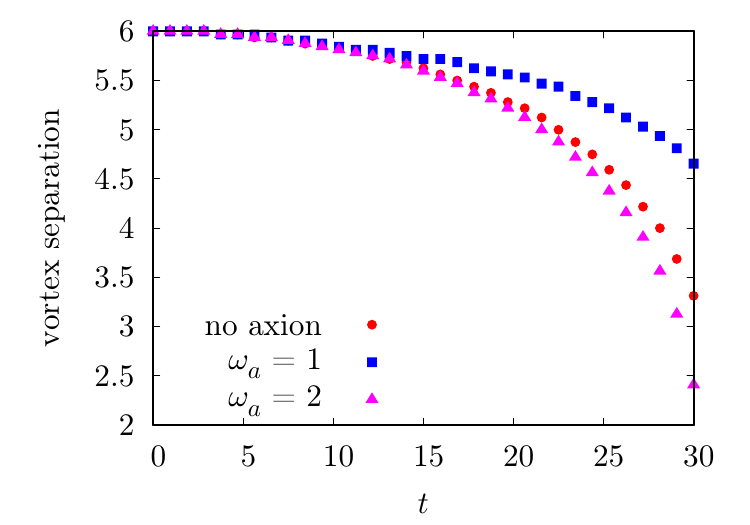}
\caption{
Time evolution of the vortex-antivortex separation. The red circles, blue squares and magenta triangles correspond respectively to the case without the axion, $\omega_a = 1$ and $\omega_a = 2$.
}
\label{fig:vortex_separation}
\end{figure}

Here we briefly discuss the vortex-antivortex interaction in the oscillating axion background, which was not studied in our previous paper \cite{Kitajima:2025nml}
but can be relevant to the dynamics of the vortex-antivortex pair creation.

We solve (\ref{Higgs_Lange}), (\ref{Ampere_Lange}) and (\ref{Gauss_Lange}) without the dissipation term and the thermal noise.
We set the initial field values describing the vortex and antivortex at rest, which are placed with some separation.
Specifically, we use the superposition ansatz for the initial field values \cite{Vilenkin:2000jqa},
\begin{align}
    \Phi(\bm{x}) &= \Phi^{(+)}(|\bm{x}-\bm{x}_1|) \Phi^{(-)}(|\bm{x}-\bm{x}_2|),\\
    A_i(\bm{x}) &= A_{i}^{(+)}(|\bm{x}-\bm{x}_1|) + A_{i}^{(-)}(|\bm{x}-\bm{x}_2|).
\end{align}
Here $\Phi^{(s)}(|\bm{x}-\bm{x}_c|)$ and $A_i^{(s)}(|\bm{x}-\bm{x}_c|)$ ($s=\pm$) are single vortex profiles with the winding number $\pm1$ for the Higgs and the gauge fields, centered at $\bm{x} = \bm{x}_c$.
The vortex profile is given by the static ANO solution in the absence of the axion background.
We performed the lattice simulation with the grid number $N^2 = 4096^2$, the boxsize $L = 64/v$ and the stepsize for time evolution $0.2 \delta x$ with $\delta x = L/N$ the lattice spacing.

\FIG{fig:vortex_separation} shows the time evolution of the separation $d$ of the vortex and antivortex from the initial separation $d_{\rm ini}=6$.
For the case of the two vortices, one can refer to Fig.~3 in \cite{Kitajima:2025nml}.
The blue squares and the magenta triangles denote the results for $\omega_a=1,2$, respectively, while the red circles correspond to the case without axion. 
In the case of two vortices, we find that the interaction is attractive for $\omega_a=1$, while it is repulsive for $\omega_a=2$ \cite{Kitajima:2025nml}.
In the case with vortex and antivortex, one can see the opposite behavior. That is, the attraction becomes weaker (stronger) in the case with $\omega_a = 1$ ($2$) compared to the case without axion.
In any case, the oscillating axion background can have a significant effect on the vortex-antivortex interaction.
However, it should rely on the complicated configurations and their time-evolution, which requires more detailed analysis and will be clarified in future work.

\section{Constraint-preserving noise
\label{app:Gauss}}

In this Appendix, we discuss the Gauss constraint in the presence of interaction with the external thermal bath.
The addition of random noise generally violates the Gauss constraint (\ref{Gauss_Lange}), unless the noise is introduced in a proper way.
Indeed, the time derivative of the quantity $G = \del_i E_i + 2e{\rm{Im}}(\Phi^*\Pi)$ is nonzero, i.e. $\dot{G} = d$, with $d$ being the residual given by
\begin{align}
d = \del_i \xi_{E_i} + e(\phi_1\xi_{\pi_2} -\phi_2 \xi_{\pi_1}),
\label{Gauss_perturb}
\end{align}
which implies that the Gauss constraint, $G=0$, gets violated even if it is initially satisfied.
Thus, we need to choose the fluctuations $\xi_\alpha$ that maintain the Gauss's law.
This problem has been studied for a long time for simulations in condensed matter physics as well as particle physics \cite{Grigoriev:1989je,Ambjorn:1990pu,Ambjorn:1990wn,Krasnitz:1992ja,Krasnitz:1993np,Krasnitz:1995xi,Gleiser:1997vb,Bettencourt:1999kk}.

Let us define the momentum vector, $q = (\vec{E},\pi_1,\pi_2)^T$, and the noise vector, $\xi = (\xi_{\vec{E}},\xi_{\pi_1},\xi_{\pi_2})^T$ and introduce an operator, $C = (\vec \del,-e\phi_2,e\phi_1)$.
Then, the Gauss constraint and the residual can be formally expressed as $Cq = 0$ and $C\xi = d$.
Let us further introduce an auxiliary field, $\Lambda$, satisfying
\begin{align}
(C C^\dagger) \Lambda = d,
\end{align}
where $C^\dagger = (-\vec \del,-e\phi_2,e\phi_1)^T$ is the conjugate operator of $C$.
Then, the projected noise vector defined by
\begin{align}
\tilde{\xi} = \xi -C^\dagger \Lambda \equiv P \xi
\end{align}
satisfies $C \tilde\xi = 0$.
This means that, by replacing the original noise vector with the projected one, the Gauss constraint is preserved.
The projection operator is obtained by solving the Poisson-like equation for $\Lambda$,
\begin{align}
(-\nabla^2 +2e^2|\Phi|^2) \Lambda = \del_i \xi_{Ei} + 2e{\rm{Im}}(\Phi^*\xi_\Pi).
\label{Gauss_preserved_app}
\end{align}

The projection operator can be algebraically expressed as
\begin{align}
P = I -C^\dagger(CC^\dagger)^{-1}C,
\end{align}
with $I$ being the identity operator.
Note that $P$ is self-conjugate, $P^\dagger = P$, and satisfies $P^2 = P$.
Then, the correlation function for $\tilde{\xi}$ reads
\begin{align}
\langle\tilde{\xi}_\alpha(t,\vec x) \tilde{\xi}_\beta(t',\vec x')\rangle &= 
P_{\alpha\gamma} P_{\beta\delta} \langle\xi_\gamma(t,\vec x) \xi_\delta(t',\vec x')\rangle \nonumber\\
&=2\eta T P_{\alpha\beta}\delta(t-t')\delta^2(\vec x-\vec x').
\end{align}
Although $P$ is inserted in the right-hand side of the fluctuation-dissipation relation (\ref{FDT}), the viscosity term in the Langevin equations does not change, thanks to $Pq = q$.
In our numerical simulations, we solve the algebraic equation (\ref{Gauss_preserved_app}) at every step using the iterative method and substitute the projected noise in the Langevin equations (\ref{Higgs_Lange}), (\ref{Ampere_Lange}).

\bibliography{reference, nonlisted}

@misc{Tong:SFT,
    author = {Tong, David},
    title = {Statistical field Theory},
    howpublished = {Lecture notes},
    year = {2017},
    url          = {https://davidtong.org/teaching/statistical-field-theory/},
}

@article{Kosterlitz_2016,
doi = {10.1088/0034-4885/79/2/026001},
url = {https://doi.org/10.1088/0034-4885/79/2/026001},
year = {2016},
month = {jan},
publisher = {IOP Publishing},
volume = {79},
number = {2},
pages = {026001},
author = {Kosterlitz, J Michael},
title = {Kosterlitz–Thouless physics: a review of key issues},
journal = {Reports on Progress in Physics}
}

@article{PhysRevLett.131.186002,
  title = {Sharpness of the Berezinskii-Kosterlitz-Thouless Transition in Disordered NbN Films},
  author = {Weitzel, Alexander and Pfaffinger, Lea and Maccari, Ilaria and Kronfeldner, Klaus and Huber, Thomas and Fuchs, Lorenz and Mallord, James and Linzen, Sven and Il'ichev, Evgeni and Paradiso, Nicola and Strunk, Christoph},
  journal = {Phys. Rev. Lett.},
  volume = {131},
  issue = {18},
  pages = {186002},
  numpages = {6},
  year = {2023},
  month = {Nov},
  publisher = {American Physical Society},
  doi = {10.1103/PhysRevLett.131.186002},
  url = {https://link.aps.org/doi/10.1103/PhysRevLett.131.186002}
}

@article{PhysRevLett.42.1165,
  title = {Possibility of Vortex-Antivortex Pair Dissociation in Two-Dimensional Superconductors},
  author = {Beasley, M. R. and Mooij, J. E. and Orlando, T. P.},
  journal = {Phys. Rev. Lett.},
  volume = {42},
  issue = {17},
  pages = {1165--1168},
  numpages = {0},
  year = {1979},
  month = {Apr},
  publisher = {American Physical Society},
  doi = {10.1103/PhysRevLett.42.1165},
  url = {https://link.aps.org/doi/10.1103/PhysRevLett.42.1165}
}

@article{halperin1979resistive,
  title={Resistive transition in superconducting films},
  author={Halperin, BI and Nelson, David R},
  journal={Journal of low temperature physics},
  volume={36},
  number={5},
  pages={599--616},
  year={1979},
  publisher={Springer}
}

@article{PhysRevB.28.5075,
  title = {Superconducting phase transitions in indium/indium-oxide thin-film composites},
  author = {Fiory, A. T. and Hebard, A. F. and Glaberson, W. I.},
  journal = {Phys. Rev. B},
  volume = {28},
  issue = {9},
  pages = {5075--5087},
  numpages = {0},
  year = {1983},
  month = {Nov},
  publisher = {American Physical Society},
  doi = {10.1103/PhysRevB.28.5075},
  url = {https://link.aps.org/doi/10.1103/PhysRevB.28.5075}
}

@article{jelic2011quench,
  title={Quench dynamics of the 2d XY model},
  author={Jeli{\'c}, Asja and Cugliandolo, Leticia F},
  journal={Journal of Statistical Mechanics: Theory and Experiment},
  volume={2011},
  number={02},
  pages={P02032},
  year={2011}
}

@article{Peccei:1977hh,
    author = "Peccei, R. D. and Quinn, Helen R.",
    title = "{CP Conservation in the Presence of Instantons}",
    reportNumber = "ITP-568-STANFORD",
    doi = "10.1103/PhysRevLett.38.1440",
    journal = "Phys. Rev. Lett.",
    volume = "38",
    pages = "1440--1443",
    year = "1977"
}

@article{Peccei:1977ur,
    author = "Peccei, R. D. and Quinn, Helen R.",
    title = "{Constraints Imposed by CP Conservation in the Presence of Instantons}",
    reportNumber = "ITP-572-STANFORD",
    doi = "10.1103/PhysRevD.16.1791",
    journal = "Phys. Rev. D",
    volume = "16",
    pages = "1791--1797",
    year = "1977"
}

@article{Weinberg:1977ma,
    author = "Weinberg, Steven",
    title = "{A New Light Boson?}",
    reportNumber = "HUTP-77/A074",
    doi = "10.1103/PhysRevLett.40.223",
    journal = "Phys. Rev. Lett.",
    volume = "40",
    pages = "223--226",
    year = "1978"
}

@article{Wilczek:1977pj,
    author = "Wilczek, Frank",
    title = "{Problem of Strong  $P$  and  $T$  Invariance in the Presence of Instantons}",
    reportNumber = "Print-77-0939 (COLUMBIA)",
    doi = "10.1103/PhysRevLett.40.279",
    journal = "Phys. Rev. Lett.",
    volume = "40",
    pages = "279--282",
    year = "1978"
}

@article{Svrcek:2006yi,
    author = "Svrcek, Peter and Witten, Edward",
    title = "{Axions In String Theory}",
    eprint = "hep-th/0605206",
    archivePrefix = "arXiv",
    reportNumber = "SLAC-PUB-11894",
    doi = "10.1088/1126-6708/2006/06/051",
    journal = "JHEP",
    volume = "06",
    pages = "051",
    year = "2006"
}

@article{Arvanitaki:2009fg,
    author = "Arvanitaki, Asimina and Dimopoulos, Savas and Dubovsky, Sergei and Kaloper, Nemanja and March-Russell, John",
    title = "{String Axiverse}",
    eprint = "0905.4720",
    archivePrefix = "arXiv",
    primaryClass = "hep-th",
    doi = "10.1103/PhysRevD.81.123530",
    journal = "Phys. Rev. D",
    volume = "81",
    pages = "123530",
    year = "2010"
}

@article{Cicoli:2012sz,
    author = "Cicoli, Michele and Goodsell, Mark and Ringwald, Andreas",
    title = "{The type IIB string axiverse and its low-energy phenomenology}",
    eprint = "1206.0819",
    archivePrefix = "arXiv",
    primaryClass = "hep-th",
    reportNumber = "DESY-12-058, CERN-PH-TH-2012-153",
    doi = "10.1007/JHEP10(2012)146",
    journal = "JHEP",
    volume = "10",
    pages = "146",
    year = "2012"
}

@article{Preskill:1982cy,
    author = "Preskill, John and Wise, Mark B. and Wilczek, Frank",
    editor = "Srednicki, M. A.",
    title = "{Cosmology of the Invisible Axion}",
    reportNumber = "HUTP-82-A048, NSF-ITP-82-103",
    doi = "10.1016/0370-2693(83)90637-8",
    journal = "Phys. Lett. B",
    volume = "120",
    pages = "127--132",
    year = "1983"
}

@article{Abbott:1982af,
    author = "Abbott, L. F. and Sikivie, P.",
    editor = "Srednicki, M. A.",
    title = "{A Cosmological Bound on the Invisible Axion}",
    reportNumber = "PRINT-82-0695 (BRANDEIS)",
    doi = "10.1016/0370-2693(83)90638-X",
    journal = "Phys. Lett. B",
    volume = "120",
    pages = "133--136",
    year = "1983"
}

@article{Dine:1982ah,
    author = "Dine, Michael and Fischler, Willy",
    editor = "Srednicki, M. A.",
    title = "{The Not So Harmless Axion}",
    reportNumber = "UPR-0201T",
    doi = "10.1016/0370-2693(83)90639-1",
    journal = "Phys. Lett. B",
    volume = "120",
    pages = "137--141",
    year = "1983"
}

@article{Sikivie:2006ni,
    author = "Sikivie, Pierre",
    editor = "Kuster, Markus and Raffelt, Georg and Beltran, Berta",
    title = "{Axion Cosmology}",
    eprint = "astro-ph/0610440",
    archivePrefix = "arXiv",
    reportNumber = "UFIFT-HEP-06-16",
    doi = "10.1007/978-3-540-73518-2_2",
    journal = "Lect. Notes Phys.",
    volume = "741",
    pages = "19--50",
    year = "2008"
}

@article{Kawasaki:2013ae,
    author = "Kawasaki, Masahiro and Nakayama, Kazunori",
    title = "{Axions: Theory and Cosmological Role}",
    eprint = "1301.1123",
    archivePrefix = "arXiv",
    primaryClass = "hep-ph",
    reportNumber = "ICRR-REPORT-643-2012-32, IPMU13-0009, UT-13-01",
    doi = "10.1146/annurev-nucl-102212-170536",
    journal = "Ann. Rev. Nucl. Part. Sci.",
    volume = "63",
    pages = "69--95",
    year = "2013"
}

@article{Marsh:2015xka,
    author = "Marsh, David J. E.",
    title = "{Axion Cosmology}",
    eprint = "1510.07633",
    archivePrefix = "arXiv",
    primaryClass = "astro-ph.CO",
    reportNumber = "KCL-PH-TH-2015-50",
    doi = "10.1016/j.physrep.2016.06.005",
    journal = "Phys. Rept.",
    volume = "643",
    pages = "1--79",
    year = "2016"
}

@article{DiLuzio:2020wdo,
    author = "Di Luzio, Luca and Giannotti, Maurizio and Nardi, Enrico and Visinelli, Luca",
    title = "{The landscape of QCD axion models}",
    eprint = "2003.01100",
    archivePrefix = "arXiv",
    primaryClass = "hep-ph",
    reportNumber = "DESY 20-036, DESY-20-036",
    doi = "10.1016/j.physrep.2020.06.002",
    journal = "Phys. Rept.",
    volume = "870",
    pages = "1--117",
    year = "2020"
}

@article{Choi:2020rgn,
    author = "Choi, Kiwoon and Im, Sang Hui and Shin, Chang Sub",
    title = "{Recent Progress in the Physics of Axions and Axion-Like Particles}",
    eprint = "2012.05029",
    archivePrefix = "arXiv",
    primaryClass = "hep-ph",
    reportNumber = "CTPU-PTC-20-28",
    doi = "10.1146/annurev-nucl-120720-031147",
    journal = "Ann. Rev. Nucl. Part. Sci.",
    volume = "71",
    pages = "225--252",
    year = "2021"
}

@article{OHare:2024nmr,
    author = "O'Hare, Ciaran A. J.",
    title = "{Cosmology of axion dark matter}",
    eprint = "2403.17697",
    archivePrefix = "arXiv",
    primaryClass = "hep-ph",
    doi = "10.22323/1.454.0040",
    journal = "PoS",
    volume = "COSMICWISPers",
    pages = "040",
    year = "2024"
}

@article{Irastorza:2018dyq,
    author = "Irastorza, Igor G. and Redondo, Javier",
    title = "{New experimental approaches in the search for axion-like particles}",
    eprint = "1801.08127",
    archivePrefix = "arXiv",
    primaryClass = "hep-ph",
    doi = "10.1016/j.ppnp.2018.05.003",
    journal = "Prog. Part. Nucl. Phys.",
    volume = "102",
    pages = "89--159",
    year = "2018"
}

@article{Abrikosov:1957wnz,
    author = "Abrikosov, A. A.",
    title = "{The magnetic properties of superconducting alloys}",
    doi = "10.1016/0022-3697(57)90083-5",
    journal = "J. Phys. Chem. Solids",
    volume = "2",
    number = "3",
    pages = "199--208",
    year = "1957"
}

@article{Nielsen:1973cs,
    author = "Nielsen, Holger Bech and Olesen, P.",
    editor = "Taylor, J. C.",
    title = "{Vortex Line Models for Dual Strings}",
    doi = "10.1016/0550-3213(73)90350-7",
    journal = "Nucl. Phys. B",
    volume = "61",
    pages = "45--61",
    year = "1973"
}

@article{Kitajima:2025nml,
    author = "Kitajima, Naoya and Nakagawa, Shota",
    title = "{Abelian-Higgs vortices in the oscillating axion background}",
    eprint = "2507.16720",
    archivePrefix = "arXiv",
    primaryClass = "hep-ph",
    doi = "10.1016/j.physletb.2025.139993",
    journal = "Phys. Lett. B",
    volume = "871",
    pages = "139993",
    year = "2025"
}

@article{Berezinskii:1972fet,
    author = "Berezinskii, V. . L.",
    title = "{Destruction of Long-range Order in One-dimensional and Two-dimensional Systems Possessing a Continuous Symmetry Group. II. Quantum Systems}",
    journal = "Sov. Phys. JETP",
    volume = "34",
    number = "3",
    pages = "610--616",
    year = "1972"
}

@article{Kosterlitz:1973xp,
    author = "Kosterlitz, J. M. and Thouless, D. J.",
    title = "{Ordering, metastability and phase transitions in two-dimensional systems}",
    doi = "10.1088/0022-3719/6/7/010",
    journal = "J. Phys. C",
    volume = "6",
    pages = "1181--1203",
    year = "1973"
}

@article{Kosterlitz:1974nba,
    author = "Kosterlitz, J. M.",
    title = "{The critical properties of the two-dimensional xy model}",
    doi = "10.1088/0022-3719/7/6/005",
    journal = "J. Phys. C",
    volume = "7",
    number = "6",
    pages = "1046",
    year = "1974"
}

@article{Nelson:1977zz,
    author = "Nelson, David R. and Kosterlitz, J. M.",
    title = "{Universal Jump in the Superfluid Density of Two-Dimensional Superfluids}",
    doi = "10.1103/PhysRevLett.39.1201",
    journal = "Phys. Rev. Lett.",
    volume = "39",
    pages = "1201--1205",
    year = "1977"
}

@article{Kitajima:2021bjq,
    author = "Kitajima, Naoya and Nakagawa, Shota and Takahashi, Fuminobu",
    title = "{Nonthermally trapped inflation by tachyonic dark photon production}",
    eprint = "2111.06696",
    archivePrefix = "arXiv",
    primaryClass = "hep-ph",
    reportNumber = "TU-1131",
    doi = "10.1103/PhysRevD.105.103011",
    journal = "Phys. Rev. D",
    volume = "105",
    number = "10",
    pages = "103011",
    year = "2022"
}

@article{Nakagawa:2022knn,
    author = "Nakagawa, Shota and Takahashi, Fuminobu and Yin, Wen",
    title = "{Early dark energy by a dark Higgs field and axion-induced nonthermal trapping}",
    eprint = "2209.01107",
    archivePrefix = "arXiv",
    primaryClass = "astro-ph.CO",
    reportNumber = "TU-1167",
    doi = "10.1103/PhysRevD.107.063016",
    journal = "Phys. Rev. D",
    volume = "107",
    number = "6",
    pages = "063016",
    year = "2023"
}

@article{Kitajima:2025jct,
    author = "Kitajima, Naoya and Uwabo-Niibo, Michiru",
    title = "{Multimodal axion emissions from Abelian-Higgs cosmic strings}",
    eprint = "2510.10708",
    archivePrefix = "arXiv",
    primaryClass = "hep-ph",
    reportNumber = "TU-1283",
    doi = "10.1103/6m7q-qf1p",
    journal = "Phys. Rev. D",
    volume = "113",
    number = "10",
    pages = "103528",
    year = "2026"
}

@article{Davis:1986xc,
    author = "Davis, Richard Lynn",
    title = "{Cosmic Axions from Cosmic Strings}",
    reportNumber = "SLAC-PUB-3895",
    doi = "10.1016/0370-2693(86)90300-X",
    journal = "Phys. Lett. B",
    volume = "180",
    pages = "225--230",
    year = "1986"
}

@article{Lyth:1991bb,
    author = "Lyth, David H.",
    title = "{Estimates of the cosmological axion density}",
    reportNumber = "LANCS-TH-17-91",
    doi = "10.1016/0370-2693(92)91590-6",
    journal = "Phys. Lett. B",
    volume = "275",
    pages = "279--283",
    year = "1992"
}

@article{Hiramatsu:2010yn,
    author = "Hiramatsu, Takashi and Kawasaki, Masahiro and Saikawa, Ken'ichi",
    title = "{Evolution of String-Wall Networks and Axionic Domain Wall Problem}",
    eprint = "1012.4558",
    archivePrefix = "arXiv",
    primaryClass = "astro-ph.CO",
    reportNumber = "ICRR-REPORT-577-2010-10, IPMU10-0221, YITP-10-110",
    doi = "10.1088/1475-7516/2011/08/030",
    journal = "JCAP",
    volume = "08",
    pages = "030",
    year = "2011"
}

@article{Hiramatsu:2010yu,
    author = "Hiramatsu, Takashi and Kawasaki, Masahiro and Sekiguchi, Toyokazu and Yamaguchi, Masahide and Yokoyama, Jun'ichi",
    title = "{Improved estimation of radiated axions from cosmological axionic strings}",
    eprint = "1012.5502",
    archivePrefix = "arXiv",
    primaryClass = "hep-ph",
    reportNumber = "IPMU10-0229, RESCEU-29-10, YITP-10-111",
    doi = "10.1103/PhysRevD.83.123531",
    journal = "Phys. Rev. D",
    volume = "83",
    pages = "123531",
    year = "2011"
}

@article{Hiramatsu:2012gg,
    author = "Hiramatsu, Takashi and Kawasaki, Masahiro and Saikawa, Ken'ichi and Sekiguchi, Toyokazu",
    title = "{Production of dark matter axions from collapse of string-wall systems}",
    eprint = "1202.5851",
    archivePrefix = "arXiv",
    primaryClass = "hep-ph",
    reportNumber = "ICRR-REPORT-608-2011-25, IPMU12-0025, YITP-12-9",
    doi = "10.1103/PhysRevD.85.105020",
    journal = "Phys. Rev. D",
    volume = "85",
    pages = "105020",
    year = "2012",
    note = "[Erratum: Phys.Rev.D 86, 089902 (2012)]"
}

@article{Hiramatsu:2012sc,
    author = "Hiramatsu, Takashi and Kawasaki, Masahiro and Saikawa, Ken'ichi and Sekiguchi, Toyokazu",
    title = "{Axion cosmology with long-lived domain walls}",
    eprint = "1207.3166",
    archivePrefix = "arXiv",
    primaryClass = "hep-ph",
    reportNumber = "ICRR-REPORT-620-2012-9, IPMU12-0140, YITP-12-58",
    doi = "10.1088/1475-7516/2013/01/001",
    journal = "JCAP",
    volume = "01",
    pages = "001",
    year = "2013"
}

@article{Kawasaki:2014sqa,
    author = "Kawasaki, Masahiro and Saikawa, Ken'ichi and Sekiguchi, Toyokazu",
    title = "{Axion dark matter from topological defects}",
    eprint = "1412.0789",
    archivePrefix = "arXiv",
    primaryClass = "hep-ph",
    reportNumber = "ICRR-REPORT-696-2014-22, IPMU14-0348",
    doi = "10.1103/PhysRevD.91.065014",
    journal = "Phys. Rev. D",
    volume = "91",
    number = "6",
    pages = "065014",
    year = "2015"
}

@article{Jacobs:1978ch,
    author = "Jacobs, Laurence and Rebbi, Claudio",
    title = "{Interaction Energy of Superconducting Vortices}",
    reportNumber = "Print-78-1156 (BNL), BNL-25246",
    doi = "10.1103/PhysRevB.19.4486",
    journal = "Phys. Rev. B",
    volume = "19",
    pages = "4486--4494",
    year = "1979"
}

@article{Taubes:1979tm,
    author = "Taubes, Clifford Henry",
    title = "{Arbitrary N: Vortex Solutions to the First Order Landau-Ginzburg Equations}",
    reportNumber = "HUTMP 79/B73",
    doi = "10.1007/BF01197552",
    journal = "Commun. Math. Phys.",
    volume = "72",
    pages = "277--292",
    year = "1980"
}

@article{Ruback:1988ba,
    author = "Ruback, P. J.",
    title = "{Vortex String Motion in the Abelian Higgs Model}",
    doi = "10.1016/0550-3213(88)90038-7",
    journal = "Nucl. Phys. B",
    volume = "296",
    pages = "669--678",
    year = "1988"
}

@article{Moriarty:1988fx,
    author = "Moriarty, K. J. M. and Myers, Eric and Rebbi, Claudio",
    title = "{Dynamical Interactions of Flux Vortices in Superconductors}",
    reportNumber = "BUHEP-88-6",
    doi = "10.1016/0370-2693(88)90674-0",
    journal = "Phys. Lett. B",
    volume = "207",
    pages = "411--418",
    year = "1988"
}

@article{Myers:1991yh,
    author = "Myers, Eric and Rebbi, Claudio and Strilka, Richard",
    title = "{A Study of the interaction and scattering of vortices in the Abelian Higgs (or Ginzburg-Landau) model}",
    reportNumber = "BUHEP-91-7",
    doi = "10.1103/PhysRevD.45.1355",
    journal = "Phys. Rev. D",
    volume = "45",
    pages = "1355--1364",
    year = "1992"
}

@article{Samols:1991ne,
    author = "Samols, T. M.",
    title = "{Vortex scattering}",
    reportNumber = "DAMTP-91-13",
    doi = "10.1007/BF02099284",
    journal = "Commun. Math. Phys.",
    volume = "145",
    pages = "149--180",
    year = "1992"
}

@article{Rebbi:1991wk,
    author = "Rebbi, C. and Strilka, R. and Myers, E.",
    title = "{Vortex scattering in the Abelian Higgs model}",
    doi = "10.1016/0920-5632(92)90243-L",
    journal = "Nucl. Phys. B Proc. Suppl.",
    volume = "26",
    pages = "240--246",
    year = "1992"
}

@article{Speight:1996px,
    author = "Speight, J. M.",
    title = "{Static intervortex forces}",
    eprint = "hep-th/9603155",
    archivePrefix = "arXiv",
    reportNumber = "PRINT-97-030 (TEXAS)",
    doi = "10.1103/PhysRevD.55.3830",
    journal = "Phys. Rev. D",
    volume = "55",
    pages = "3830--3835",
    year = "1997"
}

@article{Eto:2022hyt,
    author = "Eto, Minoru and Hamada, Yu and Jinno, Ryusuke and Nitta, Muneto and Yamada, Masatoshi",
    title = "{Abrikosov-Nielsen-Olesen strings from the Coleman-Weinberg potential}",
    eprint = "2205.04394",
    archivePrefix = "arXiv",
    primaryClass = "hep-ph",
    reportNumber = "YGHP-22-01, KEK-TH-2418, IFT-UAM/CSIC-22-48, DESY-22-071",
    doi = "10.1103/PhysRevD.106.116002",
    journal = "Phys. Rev. D",
    volume = "106",
    number = "11",
    pages = "116002",
    year = "2022"
}

@article{Fujikura:2023lil,
    author = "Fujikura, Kohei and Li, Siyao and Yamaguchi, Masahide",
    title = "{Interactions between several types of cosmic strings}",
    eprint = "2309.05515",
    archivePrefix = "arXiv",
    primaryClass = "hep-ph",
    reportNumber = "UT-Komaba/23-11",
    doi = "10.1007/JHEP12(2023)115",
    journal = "JHEP",
    volume = "12",
    pages = "115",
    year = "2023"
}

@article{Hirose:2024wuk,
    author = "Hirose, Takuya and Kanda, Yukihiro",
    title = "{Distance-dependent interaction between cosmic strings inspired by higher-dimensional gauge theory}",
    eprint = "2409.18754",
    archivePrefix = "arXiv",
    primaryClass = "hep-ph",
    doi = "10.1007/JHEP12(2024)030",
    journal = "JHEP",
    volume = "12",
    pages = "030",
    year = "2024"
}

@article{Bogomolny:1975de,
    author = "Bogomolny, E. B.",
    title = "{Stability of Classical Solutions}",
    reportNumber = "PRINT-76-0543 (LANDAU-INST.)",
    journal = "Sov. J. Nucl. Phys.",
    volume = "24",
    pages = "449",
    year = "1976"
}

@article{Prasad:1975kr,
    author = "Prasad, M. K. and Sommerfield, Charles M.",
    title = "{An Exact Classical Solution for the 't Hooft Monopole and the Julia-Zee Dyon}",
    reportNumber = "COO-3075-114",
    doi = "10.1103/PhysRevLett.35.760",
    journal = "Phys. Rev. Lett.",
    volume = "35",
    pages = "760--762",
    year = "1975"
}

@article{Mermin:1966fe,
    author = "Mermin, N. D. and Wagner, H.",
    title = "{Absence of ferromagnetism or antiferromagnetism in one-dimensional or two-dimensional isotropic Heisenberg models}",
    doi = "10.1103/PhysRevLett.17.1133",
    journal = "Phys. Rev. Lett.",
    volume = "17",
    pages = "1133--1136",
    year = "1966"
}

@article{Minnhagen:1987zz,
    author = "Minnhagen, Petter",
    title = "{The two-dimensional Coulomb gas, vortex unbinding, and superfluid-superconducting films}",
    doi = "10.1103/RevModPhys.59.1001",
    journal = "Rev. Mod. Phys.",
    volume = "59",
    pages = "1001--1066",
    year = "1987"
}

@inproceedings{Drouin-Touchette:2022ydh,
    author = "Drouin-Touchette, Victor",
    title = "{The Kosterlitz-Thouless phase transition: an introduction for the intrepid student}",
    eprint = "2207.13748",
    archivePrefix = "arXiv",
    primaryClass = "cond-mat.stat-mech",
    month = "7",
    year = "2022"
}

@article{Yates:1998kx,
    author = "Yates, Andrew and Zurek, Wojciech H.",
    title = "{Vortex formation in two-dimensions: When symmetry breaks, how big are the pieces?}",
    eprint = "hep-ph/9801223",
    archivePrefix = "arXiv",
    doi = "10.1103/PhysRevLett.80.5477",
    journal = "Phys. Rev. Lett.",
    volume = "80",
    pages = "5477--5480",
    year = "1998"
}

@article{Bettencourt:2001ru,
    author = "Bettencourt, Luis M. A. and Stephens, Greg J.",
    title = "{A Vortex description of the first order phase transition in type I superconductors}",
    eprint = "cond-mat/0108125",
    archivePrefix = "arXiv",
    reportNumber = "MIT-CTP-3169, LAUR-01-4286",
    doi = "10.1103/PhysRevE.67.066105",
    journal = "Phys. Rev. E",
    volume = "67",
    pages = "066105",
    year = "2003"
}

@article{Kajantie:1998bg,
    author = "Kajantie, K. and Karjalainen, M. and Laine, M. and Peisa, J. and Rajantie, A.",
    title = "{Thermodynamics of gauge invariant U(1) vortices from lattice Monte Carlo simulations}",
    eprint = "hep-ph/9803367",
    archivePrefix = "arXiv",
    reportNumber = "CERN-TH-98-70, HIP-1998-14-TH",
    doi = "10.1016/S0370-2693(98)00440-7",
    journal = "Phys. Lett. B",
    volume = "428",
    pages = "334--341",
    year = "1998"
}

@article{Gasenzer:2013era,
    author = "Gasenzer, Thomas and McLerran, Larry and Pawlowski, Jan M. and Sexty, D{\'e}nes",
    title = "{Gauge turbulence, topological defect dynamics, and condensation in Higgs models}",
    eprint = "1307.5301",
    archivePrefix = "arXiv",
    primaryClass = "hep-ph",
    doi = "10.1016/j.nuclphysa.2014.07.030",
    journal = "Nucl. Phys. A",
    volume = "930",
    pages = "163--186",
    year = "2014"
}

@book{Vilenkin:2000jqa,
    author = "Vilenkin, A. and Shellard, E. P. S.",
    title = "{Cosmic Strings and Other Topological Defects}",
    isbn = "978-0-521-65476-0",
    publisher = "Cambridge University Press",
    month = "7",
    year = "2000"
}

@article{Grigoriev:1989je,
    author = "Grigoriev, Dmitri Yu. and Rubakov, V. A. and Shaposhnikov, M. E.",
    title = "{Sphaleron Transitions at Finite Temperatures: Numerical Study in (1+1)-dimensions}",
    doi = "10.1016/0370-2693(89)91390-7",
    journal = "Phys. Lett. B",
    volume = "216",
    pages = "172--176",
    year = "1989"
}

@article{Ambjorn:1990pu,
    author = "Ambjorn, Jan and Askgaard, T. and Porter, H. and Shaposhnikov, M. E.",
    title = "{Sphaleron transitions and baryon asymmetry: A Numerical real time analysis}",
    reportNumber = "NBI-HE-90-48",
    doi = "10.1016/0550-3213(91)90341-T",
    journal = "Nucl. Phys. B",
    volume = "353",
    pages = "346--378",
    year = "1991"
}

@article{Ambjorn:1990wn,
    author = "Ambjorn, Jan and Askgaard, T. and Porter, H. and Shaposhnikov, M. E.",
    title = "{Lattice Simulations of Electroweak Sphaleron Transitions in Real Time}",
    reportNumber = "NBI-HE-90-18",
    doi = "10.1016/0370-2693(90)90350-F",
    journal = "Phys. Lett. B",
    volume = "244",
    pages = "479--487",
    year = "1990"
}

@article{Krasnitz:1992ja,
    author = "Krasnitz, A. and Potting, Robertus",
    title = "{A Realistic heat bath: Theory and application to kink - antikink dynamics}",
    eprint = "hep-lat/9212033",
    archivePrefix = "arXiv",
    reportNumber = "IPS-92-30, IUHET-241, UALG-PHYS-12",
    doi = "10.1016/0550-3213(93)90532-T",
    journal = "Nucl. Phys. B",
    volume = "410",
    pages = "649--662",
    year = "1993"
}

@article{Krasnitz:1993np,
    author = "Krasnitz, A. and Potting, R.",
    title = "{Sphaleron transitions in a realistic heat bath}",
    eprint = "hep-ph/9308307",
    archivePrefix = "arXiv",
    reportNumber = "IPS-93-10",
    doi = "10.1016/0370-2693(93)91545-X",
    journal = "Phys. Lett. B",
    volume = "318",
    pages = "492--500",
    year = "1993"
}

@article{Krasnitz:1995xi,
    author = "Krasnitz, A.",
    title = "{Thermalization algorithms for classical gauge theories}",
    eprint = "hep-lat/9507025",
    archivePrefix = "arXiv",
    reportNumber = "NBI-HE-95-25",
    doi = "10.1016/0550-3213(95)00465-5",
    journal = "Nucl. Phys. B",
    volume = "455",
    pages = "320--338",
    year = "1995"
}

@article{Gleiser:1997vb,
    author = "Gleiser, Marcelo and Muller, Hans-Reinhard",
    title = "{How to count kinks: From the continuum to the lattice and back}",
    eprint = "hep-lat/9704005",
    archivePrefix = "arXiv",
    reportNumber = "DART-HEP-97-04",
    doi = "10.1016/S0370-2693(97)01595-5",
    journal = "Phys. Lett. B",
    volume = "422",
    pages = "69--75",
    year = "1998"
}

@article{Bettencourt:1999kk,
    author = "Bettencourt, Luis M. A. and Habib, Salman and Lythe, Grant",
    title = "{Controlling one-dimensional Langevin dynamics on the lattice}",
    eprint = "hep-lat/9903007",
    archivePrefix = "arXiv",
    reportNumber = "LA-UR-99-891",
    doi = "10.1103/PhysRevD.60.105039",
    journal = "Phys. Rev. D",
    volume = "60",
    pages = "105039",
    year = "1999"
}

\end{document}